\documentclass{article}
\usepackage{latexsym}
\advance \topmargin by -\headheight
\advance \topmargin by -\headsep
     
\evensidemargin \oddsidemargin
\newcommand\rf[1]{(\ref{eq:#1})}
\newcommand\lab[1]{\label{eq:#1}}
\newcommand\nonu{\nonumber}
\newcommand\br{\begin{eqnarray}}
\newcommand\er{\end{eqnarray}}
\newcommand\be{\begin{equation}}
\newcommand\ee{\end{equation}}

\newcommand\foot[1]{\footnotemark\footnotetext{#1}}
\newcommand\lb{\lbrack}
\newcommand\rb{\rbrack}

\newcommand\lcurl{\left\{}
\newcommand\rcurl{\right\}}
\renewcommand\({\left(}
\renewcommand\){\right)}
\renewcommand\v{\vert}                     %% vertical bars
\newcommand\lskip{\vskip\baselineskip\vskip-\parskip\noindent}

\newcommand\bc{\begin{center}}
\newcommand\ec{\end{center}}

\newcommand\Tr{\mathop{\mathrm Tr}}                  % Tr - big trace
\newcommand\partder[2]{{{\partial {#1}}\over{\partial {#2}}}}
\newcommand\Sbr[2]{\Bigl\lbrack\,{#1}\, ,\,{#2}\,\Bigr\rbrack}
\newcommand\Pbr[2]{\Bigl\{ \,{#1}\, ,\,{#2}\,\Bigr\}}  % Poisson brackets (large)

\renewcommand\a{\alpha}
\renewcommand\b{\beta}

\renewcommand\d{\delta}

\newcommand\eps{\epsilon}
\newcommand\vareps{\varepsilon}
\newcommand\g{\gamma}
\newcommand\G{\Gamma}

\newcommand\h{{1\over 2}}
\renewcommand\k{\kappa}
\renewcommand\l{\lambda}
\renewcommand\L{\Lambda}
\newcommand\m{\mu}
\newcommand\n{\nu}
\renewcommand\o{\over}
\newcommand\om{\omega}

\newcommand\vp{\varphi}
\renewcommand\P{\Phi}
\newcommand\pa{\partial}

\newcommand\pr{\prime}

\newcommand\s{\sigma}

\renewcommand\t{\tau}
\renewcommand\th{\theta}

\newcommand\wti{\widetilde}

\newcommand\cA{{\mathcal A}}

\newcommand\cC{{\mathcal C}}
\newcommand\cD{{\mathcal D}}
\newcommand\cE{{\mathcal E}}

\newcommand\cH{{\mathcal H}}

\newcommand\cL{{\mathcal L}}
\newcommand\cM{{\mathcal M}}

\newcommand\cP{{\mathcal P}}

\newcommand\cT{{\mathcal T}}
\newcommand\cU{{\mathcal U}}

\newcommand{\ct}[1]{\cite{#1}}
\newcommand{\bi}[1]{\bibitem{#1}}
\newcommand\PRL[3]{\textsl{Phys. Rev. Lett.} \textbf{#1} (#2) #3}
\newcommand\NPB[3]{\textsl{Nucl. Phys.} \textbf{B#1} (#2) #3}

\newcommand\PRD[3]{\textsl{Phys. Rev.} \textbf{D#1} (#2) #3}

\newcommand\PLB[3]{\textsl{Phys. Lett.} \textbf{#1B} (#2) #3}
\newcommand\CQG[3]{\textsl{Class. Quantum Grav.} \textbf{#1} (#2) #3}

\newcommand\IJMPA[3]{\textsl{Int. J. Mod. Phys.} \textbf{A#1} (#2) #3}

\newcommand\MPLA[3]{\textsl{Mod. Phys. Lett.} \textbf{A#1} (#2) #3}

\begin{document}
\noindent\null\hfill {hep-th/0203024}

\begin{center}
{\Large{\textbf{String and Brane Models \\ with 
Spontaneously/Dynamically Induced Tension}}}
\end{center}    

\begin{center}
E.I. Guendelman and A. Kaganovich \\
{\small Department of Physics, Ben-Gurion University of the Negev} \\
{\small P.O.Box 653, IL-84105 ~Beer-Sheva, Israel} \\
{\small E-mail: guendel@bgumail.bgu.ac.il , alexk@bgumail.bgu.ac.il} 
\end{center}

\begin{center}
E. Nissimov and S. Pacheva  \\
{\small Institute for Nuclear Research and Nuclear Energy, 
Bulgarian Academy of Sciences} \\
{\small Boul. Tsarigradsko Chausee 72, BG-1784 ~Sofia, Bulgaria} \\
{\small E-mail: nissimov@inrne.bas.bg , svetlana@inrne.bas.bg} 
\end{center}

\begin{abstract}
We study in some detail the properties of a previously proposed new 
class of string and brane models whose world-sheet (world-volume) actions 
are built with a modified reparametrization-invariant measure of integration
and which do not contain any \textsl{ad hoc} dimensionfull parameters. The
ratio of the new and the standard Riemannian integration measure densities 
plays the role of a dynamically generated string/brane tension. The latter is 
identified as (the magnitude of) an effective (non-Abelian) electric 
field-strength on the world-sheet/world-volume obeying the standard Gauss-law
constraint. As a result a simple classical mechanism for confinement via 
modified-measure ``color'' strings is proposed where the colorlessness of the 
``hadrons'' is an automatic consequence of the new string dynamics.
\end{abstract}    
\lskip
% \section{Introduction: Main Ideas and Features of the Theory}
{\large{\textbf{1. Introduction: Main Ideas and Features of the Theory}}}

One of the characteristic features of string and brane theories
\ct{Neem-Polch} is the introduction \textsl{ad hoc}
from the very beginning of a dimensionfull scale -- the so called string
(brane) tension. On the other hand, a lot of attention has been given to 
the idea that any fundamental theory of Nature should not contain any 
\textsl{ad hoc} fundamental scales and that these scales should rather appear
as a result of dynamical generation, \textsl{e.g.}, through boundary
conditions on the classical level, spontaneous symmetry breaking and/or
dimensional transmutation on the quantum level (see, for instance,
ref.\ct{Zee-79} about spontaneous generation of Newton's gravitational
constant). 

In the context of string and brane theories, the above idea has been first
explored in refs.\ct{mstring-orig}. In this Section we will briefly review,
with some additional new accents, the main properties of the modified string
and brane theories of \ct{mstring-orig} in order to prepare the ground for
revealing of new interesting structures inherent of these theories.
To this end let us first recall the standard Polyakov-type action for the
bosonic string which reads \ct{Polyakov}:
\be
S_{\mathrm{Pol}} = - T \int d^2\s\,
\h\sqrt{-\g}\g^{ab}\pa_a X^\m \pa_b X^\n G_{\m\n}(X)
\lab{Pol-action-string}
\ee
Here $(\s^0,\s^1) \equiv (\t,\s)$; $a,b =0,1$; $\m,\n = 0,1,\ldots ,D-1$;
$G_{\m\n}$ denotes the external space-time metric;
$\g_{ab}$ is the metric defined on the $1+1$-dimensional world-sheet of 
the string and $\g = \det\v\v\g_{ab}\v\v$.
$T$ indicates the string tension -- a dimensionfull quantity introduced 
\textsl{ad hoc} into the theory which defines a scale.

Now following refs.\ct{mstring-orig}, instead of the standard measure of
integration $d^2\s\,\sqrt{-\g}$, we want to consider a new reparametrization
invariant measure on the string world-sheet whose density $\Phi$ is 
independent of the Riemannian metric $\g_{ab}$. This approach of considering 
an alternative integration measure has been studied in the context of 
$D\! =\! 4$ gravitational theory, in particular, in relation
with the cosmological constant problem \ct{GK-1} (and references 
therein), as well as the fermion families and long-range force problems \ct{GK-2}.

Indeed, if we introduce two auxiliary scalar fields (scalars both from 
the point of view of the $1+1$-dimensional world-sheet of the string, as well as 
from the point of view of the embedding $D$-dimensional universe) $\vp^{i}$ 
($i=1,2$), we can construct the following world-sheet measure density:
\be
\P (\vp) \equiv \h \vareps_{ij} \vareps^{ab} \pa_a \vp^i \pa_b \vp^j
= \vareps_{ij} {\dot{\vp}}^i \pa_\s \vp^j
\lab{def-measure}
\ee
It is interesting to notice that $d^2\s\,\P (\vp) = d\vp^1 d\vp^2$, that is 
the measure of integration $d^2\s\,\P$ corresponds to integrating in the 
target space of the auxiliary scalar fields $\vp^{i}$ ($i=1,2$).

We proceed now with the construction of a new string action that 
employs the integration measure $d^2\s\,\P$ \rf{def-measure} instead of the usual 
$d^2\s\,\sqrt{-\g}$. When considering the types of actions we can 
have under these circumstances, the first one that comes to mind is the 
straightforward generalization of the Polyakov-type action \rf{Pol-action-string} :
\be
S_{1} = -\h \int d^2\s\,\P (\vp) \g^{ab} \pa_a X^\m \pa_b X^\n G_{\m\n}(X)
\lab{S-1-def}
\ee
Notice that multiplying $S_{1}$ by a constant, before boundary or 
initial conditions are specified, is a meaningless operation since such a 
constant can be absorbed in a redefinition of the measure fields $\vp^{i}$ 
($i=1,2$) that appear in $\P (\vp)$ \rf{def-measure}.

The form \rf{S-1-def} is, however, not a satisfactory choice for a string
action because the variation of $S_{1}$ with respect to $\g^{ab}$ 
leads to the rather strong condition:
\be
\P (\vp)\, \pa_a X^\m \pa_b X^\n G_{\m\n}(X) = 0  
\lab{strong-cond}
\ee
If $\P \neq 0 $, it means that $\pa_a X^\m \pa_b X^\n G_{\m\n}(X) = 0$,
\textsl{i.e.}, it means that the metric induced on the string world-sheet
vanishes which is clearly not an acceptable dynamics. Alternatively, if 
$\P=0$, no further information is available -- also an undesirable situation.

The situation may be improved by introducing external antisymmetric tensor 
gauge field  $B_{\m\n}(X)$. Then, instead of \rf{S-1-def}, we have to 
consider the action:
\be
S_{2} = - \int d^2\s \,\P (\vp) \Bigl\lb \h \g^{ab} \pa_a X^{\m} \pa_b X^{\n}
G_{\m\n}(X) + 
\frac{\vareps^{ab}}{2\sqrt{-\g}} \pa_a X^{\m} \pa_b X^{\n} B_{\m\n}(X)
\Bigr\rb 
\lab{action-string-external}
\ee
where $\vareps^{01} = - \vareps^{10} = 1$ and $\vareps^{00} = \vareps^{11} = 0$.
Varying \rf{action-string-external} with respect to $\g^{ab}$, we get
(if $\P \neq 0 $) : 
\be
\pa_a X^{\m} \pa_b X^{\n}G_{\m\n} + 
\g_{ab}\frac{\vareps^{cd}}{4\sqrt{-\g}} \pa_c X^{\m} \pa_d X^{\n} B_{\m\n}
= 0
\lab{eq1-string-external-a}
\ee
Contracting the latter equation with $\g^{ab}$ we see that:
\be
\frac{\vareps^{cd}}{2\sqrt{-\g}} \pa_c X^{\m} \pa_d X^{\n} B_{\m\n} 
= -\g^{ab} \pa_a X^{\m} \pa_b X^{\n} G_{\m\n}
\lab{eq1-string-external-b}
\ee
Inserting relation \rf{eq1-string-external-b} into 
Eq.\rf{eq1-string-external-a} we obtain:
\be
\pa_a X^{\m} \pa_b X^{\n} G_{\m\n} - 
\h \g_{ab}\g^{cd} \pa_c X^{\m} \pa_d X^{\n} G_{\m\n} =0
\lab{Pol-like-eq}
\ee
which coincides with the form of the string equations of motion corresponding
to the Polyakov-type action \rf{Pol-action-string}
when the external antisymmetric tensor gauge field $B_{\m\n}$ is absent.

To make further progress and at the same time to show that one can
avoid the need of incorporation of an external field, it is important to 
notice that terms in the action of the form: 
\be
S = \int d^2\s\,\sqrt{-\g}\, L
\lab{action-gamma}
\ee
which do not contribute to the equations of motion of the standard closed string,
\textsl{i.e.}, such that $\sqrt{-\g} L$ is a total derivative, may yield
non-trivial contributions when we consider the counter-parts of 
\rf{action-gamma} of the form:
\be
S = \int d^2\s\,\P (\vp)\, L
\lab{action-Phi}
\ee  
This is so because if $\sqrt{-\g}\, L$ is a total divergence,
$\P L$ in general is not.                                                    

The above fact is indeed crucial. For example, let us consider the 
modified-measure string theory with an additional intrinsic $1+1$-dimensional
scalar curvature term:
\be
S_{curv} = - \int d^2\s \,\P (\vp) \, R
\lab{curv}
\ee
which now {\em is not a topological} term in contrast to
$ \int d^2\s \,\sqrt{-\g}\, R$ in the ordinary string
theory with the regular world-sheet integration measure. According to
refs.\ct{GK-1}, where modified-measure gravity theories in higher
dimensions $D>2$ have been explored, we know that in order to
achieve physically interesting results one has to proceed in the first
order formalism -- employing either the affine connection or the 
spin connection. In the present paper we will restrict ourselves by exploring the
spin connection formalism only. This means that the independent dynamical degrees
of freedom are:  zweibein $e_{a}^{\bar{a}}$, spin connection
$\omega_{a}^{\bar{a}\bar{b}}$, ($\bar{a}=0,1$ are tangent ``Lorentz'' indexes)
and the auxiliary scalar fields $\vp^{i}$ entering the new integration
measure density $\P (\vp)$ \rf{def-measure}.

We will use the following notations:
$\g^{ab}=e^{a}_{\bar{a}}e^{b}_{\bar{b}}\eta^{\bar{a}\bar{b}}$; the scalar
curvature of the spin conection is
$R(\omega ,e) =e^{a\bar{a}}e^{b\bar{b}}R_{\bar{a}\bar{b}ab}(\omega)$ where:
\be
R^{\bar{a}\bar{b}}_{ab}(\omega)=\partial_{a}\omega_{b}^{\bar{a}\bar{b}}
+\omega_{a}^{\bar{a}\bar{c}}\omega_{b \bar{c}}^{\bar{b}}
-(a\leftrightarrow b)
\lab{B}
\ee

Notice now that in $D=2$ :
\be
\omega_{a}^{\bar{a}\bar{b}}=\omega_{a}\vareps^{\bar{a}\bar{b}}
\label{omega}
\ee
where $\omega_{a}$ is a vector field. Therefore, we get for scalar curvature:
\be
R(\om) = \frac{\vareps^{ab}}{2\sqrt{-\g}} \(\pa_a \om_b - \pa_b 
\om_a\)
\lab{spin-curv}
\ee 
We conclude that the vector field $\omega_{a}$, as a geometrical
object associated with the spin-connection, can be treated as
an abelian gauge field $A_{a}$ living on the world-sheet.

Thus, let us consider an abelian gauge field $A_{a}$ defined on the world-sheet
of the string, in addition to the measure-density fields $\vp^i$ that appear in
$\P (\vp)$ \rf{def-measure}, the usual Riemannian metric $\g_{ab}$ and the 
string coordinates $X^\m$. We can then construct the following non-trivial
contribution to the action of the form:
\be
S_{gauge} = \h \int d^2\s\,\P (\vp) \frac{\vareps^{ab}}{\sqrt{-\g}}F_{ab}(A)
\quad ,\quad F_{ab} = \pa_{a} A_{b} - \pa_{b} A_{a}
\lab{action-gauge}
\ee
Therefore, the total action to be considered now is
$S_{\textrm{string}} = S_{2} + S_{gauge}$ reading explicitly:
\be
S_{\textrm{string}} = 
- \int d^2\s \,\P (\vp) \Bigl\lb \h \g^{ab} \pa_a X^{\m} \pa_b X^{\n} G_{\m\n}
+ \frac{\vareps^{ab}}{2\sqrt{-\g}}\( \pa_a X^{\m} \pa_b X^{\n} B_{\m\n} -
F_{ab}(A)\) \Bigr\rb \equiv - \int d^2\s \,\P (\vp) L
\lab{action-string}
\ee
The properties of this model and some of its generalizations will be 
studied in the following sections.

The action \rf{action-string} is invariant under a set of diffeomorphisms 
in the target space of the measure-density fields $\vp^{i}$ combined with a 
conformal (Weyl) transformation of the metric $\g_{ab}$, namely :  
\be
\vp^{i} \longrightarrow \vp^{\pr\, i} = \vp^{\pr\, i} (\vp) \quad 
\textrm{so ~that} \quad  \P \longrightarrow \P^{\pr} = J \P
\lab{vp-diff}
\ee
where $J = \det \Bigl\Vert \frac{\pa\vp^{\pr\, i}}{\pa\vp^j} \Bigr\Vert$ is 
the Jacobian of the transformation \rf{vp-diff}, and:
\be
\g_{ab} \longrightarrow \g^{\pr}_{ab} = J \g_{ab}
\lab{gamma-conf}
\ee
In what follows we will refer to the set of transformations 
\rf{vp-diff}--\rf{gamma-conf} as $\P$-{\em extended two-dimensional Weyl 
transformations} and, accordingly, to the action \rf{action-string} as being 
$\P$-extended Weyl-invariant. 
Notice also that the spin-curvature term (Eq.\rf{curv} with $R$ as in 
\rf{spin-curv}) is also $\P$-extended Weyl-invariant ($\P$-extended 
Weyl transformations do not affect the spin connection).

The combination $\frac{\vareps^{ab}}{\sqrt{-\g}} F_{ab}$ is a genuine 
scalar. In two dimensions it is proportional to $\sqrt{F_{ab}F^{ab}}$. 
In the non-Abelian case one can consider terms in the action of the 
form $\P\,\sqrt{\Tr (F_{ab} F^{ab})}$, the latter being 
$\P$-extended Weyl-invariant object ($\sqrt{Tr(F_{ab}F^{ab})}$ 
is also a genuine scalar). This model will be studied in Sec.5 below. 

To demonstrate some general features of the theory, we will first 
follow the Lagrangian formalism for solution of the modified-measure string 
model \rf{action-string} explored in refs.\ct{mstring-orig}.
Variation of the action \rf{action-string} with respect to $\vp^{i}$ yields 
the equations (here we set $B_{\m\n}=0$ for simplicity) :
\be
\vareps^{ab} \pa_{b} \vp_{i} \pa_{a} \Bigl(
\g^{cd} \pa_{c} X^{\m}\pa_{d} X^{\n} G_{\m\n}(X) -
\frac {\vareps^{cd}}{\sqrt{-\g}} F_{cd} \Bigr) = 0
\lab{vp-eqs-motion}
\ee
If $\det \v\v\vareps^{ab} \pa_{b} \vp_{i}\v\v \neq 0$
meaning $\P (\vp)\neq 0$, then we conclude that all the derivatives of
the quantity inside the parenthesis in Eq.\rf{vp-eqs-motion} must vanish,
\textsl{i.e.}, such quantity must equal certain constant $M$ which will 
be determined later on:
\be
\g^{cd} \pa_{c} X^{\m}\pa_{d} X^{\n} G_{\m\n}(X) -
\frac {\vareps^{cd}}{\sqrt{-\g}} F_{cd} = M
\lab{L-const}
\ee
The equations of motion of the gauge field $A_{a}$ tell us about how the string 
tension appears as an integration constant. Indeed, these equations are:
\be
\vareps^{ab} \pa_{b} \Bigl(\frac{\P (\vp)}{\sqrt{-\g}}\Bigr) = 0
\lab{A-eqs-motion}
\ee
which can be integrated to yield a {\em spontaneously induced} string tension: 
\be
\frac {\P (\vp)}{\sqrt{-\gamma}} = \textrm{const} \equiv T
\lab{string-tension}
\ee
Notice that Eq.\rf{string-tension} is perfectly consistent with the $\P$-
extended Weyl symmetry \rf{vp-diff}--\rf{gamma-conf}. Eq.\rf{L-const} on
the other hand is consistent with the $\P$-extended Weyl symmetry only if
$M = 0$. We will see in the next paragraph that the equations of motion
indeed imply that $M = 0$. In the case of higher-dimensional $p$-branes,
unlike the string case, the corresponding equations of motion will require 
a {\em non-vanishing} constant value of $M$ (cf. Eq.\rf{metrics-rel} below).

Let us turn our attention to the equations of motion derived from the
variation of \rf{action-string} with respect to $\g^{ab}$ :
\be
\P (\vp) \Bigl(\pa_{a} X^{\m}\pa_{b} X^{\n} G_{\m\n}(X)   
- \h\g_{ab} \frac{\vareps^{cd}}{\sqrt{-\g}}F_{cd}\Bigr) = 0
\lab{g-eqs-motion}
\ee
Solving the constraint Eq.\rf{L-const} for 
$\frac{\vareps^{cd}}{\sqrt{-\g}} F_{cd}$ and inserting the result back into
\rf{g-eqs-motion} we obtain (provided $\P (\vp) \neq 0$) :
\be
\Bigl( \pa_{a} X^{\m}\pa_{b} X^{\n} - 
\h \g_{ab} \g^{cd} \pa_{c} X^{\m}\pa_{d} X^{\n} \Bigr) G_{\m\n}(X)
+ \h \g_{ab} M = 0
\lab{L-const+g-eqs}
\ee
Multiplying the above equation by $\g^{ab}$ and summing over $a,b$,
we find that $M = 0$, \textsl{i.e.}, Eqs.\rf{L-const+g-eqs} with $M=0$ 
are exactly of the form of Eqs.\rf{Pol-like-eq} coming from the standard 
Polyakov-type action \rf{Pol-action-string} (recall also that it is only $M=0$
which is consistent with the $\P$-extended Weyl invariance). After 
Eq.\rf{string-tension} is used, the equations obtained from the variation of
the action \rf{action-string} with respect to $X^{\m}$ are seen to be exactly 
the same as those obtained from the usual Polyakov-type action as well.
\lskip
{\large{\textbf {2. Bosonic Strings with a Modified Measure: Canonical
Approach}}}
% \section{Bosonic Strings with a Modified Measure: Canonical Approach}

It is instructive to study the modified-measure string model 
\rf{action-string} also within the framework of the canonical Hamiltonian 
formalism. 

Before proceeding let us note that we can extend the model \rf{action-string}
by putting point-like charges on the string world-sheet which interact 
with the world-sheet gauge field $A_a$ :
\be
S = S_{\mathrm{string}} - \sum_i e_{i} \int d\t A_0 (\t,\s_{i})
\lab{action-charge-string}
\ee

For the canonical momenta of $\vp^i,\, A_1,\, X^\m$ we obtain (using the
short-hand notation $L$ from \rf{action-string}) :
\be
\pi^{\vp}_i = - \vareps_{ij} \pa_\s \vp^j L \quad ,\quad
\pi_{A_1} \equiv E = \frac{\P (\vp)}{\sqrt{-\g}}
\lab{momenta-1}
\ee
\be
\cP_\m = \P (\vp) \Bigl\lb - 
\(\g^{00}\dot{X}^\n + \g^{01}\pa_\s X^\n\) G_{\m\n} - 
\frac{1}{\sqrt{-\g}} \pa_\s X^\n B_{\m\n} \Bigr\rb
\lab{momenta-X}
\ee
Note particularly the second Eq.\rf{momenta-1} showing that the ratio of the
modified and the usual Riemannian integration-measure densities has the
physical meaning of an {\em electric field-strength} on the
world-sheet\foot{In analogy with ordinary electrodynamics/Yang-Mills 
theory the canonically conjugated momentum $\pi_{A_1} \equiv E$ of the 
space-like gauge-field component $A_1$ is by definition the electric field-
strength. However, unlike the ordinary case $E$ is now {\em not} proportional 
to $F_{01}(A)$; see also Sect.5 for the non-Abelian case.}.

We have also the following primary constraints:
\be
\pi_{A_0} = 0 \quad ,\quad \pi_{\g^{ab}} = 0 \quad ,\quad 
\pa_\s \vp^i \pi^{\vp}_i = 0
\lab{prim-constr}
\ee
where the last constraint follows directly from the first Eq.\rf{momenta-1}.
From \rf{momenta-1}--\rf{momenta-X} we can express the velocities in terms
of the canonical coordinates and momenta as follows:
\be
\dot{X}^\m \equiv \dot{X}^\m (\ldots ) = - \frac{G^{\m\n}}{\sqrt{-\g} 
\g^{00}} \Bigl(\frac{\cP_\n}{E} + \pa_\s X^\l B_{\n\l}\Bigr) - 
\frac{\g^{01}}{\g^{00}} \pa_\s X^\m
\lab{X-dot}
\ee
\br
\dot{A}_1 \equiv \dot{A}_1 (\ldots ) = \pa_\s A_0 
- \sqrt{-\g} \frac{\pi^{\vp}_2}{\pa_\s \vp_1} 
+ \dot{X}^\m (\ldots ) \pa_\s X^\n B_{\m\n}
\nonu \\
+ \sqrt{-\g} \Bigl( \h\g^{00} \dot{X}^\m (\ldots ) \dot{X}^\n (\ldots ) 
+ \g^{01} \dot{X}^\m (\ldots ) \pa_\s X^\n
+ \h \g^{11} \pa_\s X^\m \pa_\s X^\n \Bigr) G_{\m\n}
\lab{A1-dot}
\er
In Eq.\rf{A1-dot} we used the short-hand notation $\dot{X}^\m (\ldots)$
defined in \rf{X-dot}. Since the original Lagrangian $\cL$ in 
\rf{action-string} is homogeneous of first order with respect to $\dot{\vp}^i$ we
have $\pi^{\vp}_i \dot{\vp}^i -\cL =0$ and, therefore, the canonical Hamiltonian
reads:
\br
\cH = \cP_\m \dot{X}^\m (\ldots ) + E \dot{A}_1 (\ldots )  
\phantom{aaaaaaaaaaaaaaaaaaaaa}
\nonu \\
= - \frac{1}{\sqrt{-\g} \g^{00}} \h \Bigl\lb
\frac{G^{\m\n}}{E}\(\cP_\m + E \pa_\s X^{\m^\pr} B_{\m \m^\pr}\)
\(\cP_\n + E \pa_\s X^{\n^\pr} B_{\n \n^\pr}\) + E G_{\m\n}\pa_\s X^\m 
\pa_\s X^\n
\Bigr\rb
\nonu \\
+ \frac{\g^{01}}{\g^{00}} \cP_\m \pa_\s X^\m + E \pa_\s A_0
- E \sqrt{-\g} \frac{\pi^{\vp}_2}{\pa_\s \vp_1}
+ \sum_i e_{i} \d (\s - \s_{i})\, A_0
\lab{Ham-canon}
\er
where we used the expressions for the velocities as functions of the 
canonical coordinates and momenta \rf{X-dot}--\rf{A1-dot} and we also included 
the point-like charge interaction terms from \rf{action-charge-string}. 
Commuting of the canonical Hamiltonian \rf{Ham-canon} with the primary 
constraints \rf{prim-constr} leads to the following secondary constraints:
\be
\frac{\pi^{\vp}_2}{\pa_\s \vp_1} = 0 \quad ,\quad
\pa_\s E - \sum_i e_{i} \d (\s - \s_{i}) = 0
\lab{second-constr-a}
\ee
\be
\frac{G^{\m\n}}{E}\(\cP_\m + E \pa_\s X^{\m^\pr} B_{\m \m^\pr}\)
\(\cP_\n + E \pa_\s X^{\n^\pr} B_{\n \n^\pr}\) + E G_{\m\n}\pa_\s X^\m 
\pa_\s X^\n = 0
\lab{second-constr-b}
\ee
\be
\cP_\m \pa_\s X^\m \equiv \(\cP_\m + E \pa_\s X^\n B_{\m\n}\)\pa_\s X^\m  = 0
\lab{second-constr-c}
\ee
In particular, we obtain that the canonical Hamiltonian is a linear
combination of constraints only.

The Poisson algebra of the constraints can straightforwardly be 
computed. First, we observe that the last constraint in \rf{prim-constr} span
(centerless) Virasoro algebra:
\be
\Pbr{\pa_\s \vp^i \pi^{\vp}_i (\s)}{\pa_{\s^\pr} \vp^i \pi^{\vp}_i (\s^\pr)} =
2 \pa_\s \vp^i \pi^{\vp}_i (\s) \pa_\s \d (\s -\s^\pr) + 
\pa_\s \(\pa_\s \vp^i \pi^{\vp}_i\) \d (\s -\s^\pr)
\lab{Vir-alg}
\ee
The only nontrivial commutator of the latter with the rest of the
constraints is:
\be
\Pbr{\pa_\s \vp^i \pi^{\vp}_i (\s)}{\frac{\pi^{\vp}_2}{\pa_\s \vp_1}(\s^\pr)} =
- \pa_\s \Bigl(\frac{\pi^{\vp}_2}{\pa_\s \vp_1}\Bigr) \d (\s -\s^\pr)
\lab{Vir-1}
\ee
Therefore, both constraints $\pa_\s \vp^i \pi^{\vp}_i$ and 
$\frac{\pi^{\vp}_2}{\pa_\s \vp_1}$ span a closed algebra of first-class 
constraints, which implies that all auxiliary scalars $\vp^i$ entering the
modified measure \rf{def-measure} are pure-gauge degrees of freedom.

Next, we observe that the second constraint in \rf{second-constr-a} is
nothing but Gauss-law first-class constraint for the world-sheet 
Abelian gauge field, with $E$ being the corresponding electric field-strength. 
Obviously, $E$ is piece-wise constant (with respect to $\s$) on the world-sheet 
with jumps at the locations of the point-like charges:
\be
E = E_0  + \sum_i e_{i} \th (\s - \s_i) 
\lab{E-field}
\ee
Moreover, since the canonical Hamiltonian \rf{Ham-canon} does not depend
explicitly on $A_1$, $E$ is conserved (world-sheet time-independent).

Finally, the constraints \rf{second-constr-b}--\rf{second-constr-c}, or 
more properly, the linear combinations thereof:
\be
\cT_{\pm} \equiv {1\o 4} G^{\m\n}
\Bigl(\frac{\cP_\m}{E} \pm (G_{\m\k} \pm B_{\m\k})\pa_\s X^\k \Bigr)
\Bigl(\frac{\cP_\n}{E} \pm (G_{\n\l} \pm B_{\n\l})\pa_\s X^\l \Bigr)
\lab{two-Vir-constr}
\ee
span the same first-class constraint algebra of two mutually commuting
centerless Virasoro algebras as in the case of ordinary Polyakov-type string
(in the standard case $H_{\m\n\l}(B) \equiv 3 \pa_{\lb \m} B_{\n\l\rb}= 0$)
provided we identify the constant world-sheet electric field $E$ with 
the ordinary string tension $T$. 

To summarize so far, we find that the modified-measure string model
\rf{action-string} (or \rf{action-charge-string}), containing no 
\textsl{ad hoc} dimensionfull parameters, produces a {\em dynamically generated} 
effective string tension, which is equal to the ratio of the modified and usual
Riemannian integration-measure densities, and which has the physical meaning of
a world-sheet electric field strength. As a result the dynamical string tension
is (piece-wise) constant along the string with possible jumps at the locations 
of attached point-like charges (see Sect.5.3 for explicit examples).
\lskip
{\large{\textbf{3. Bosonic Branes with a Modified Measure}}}
% \section{Bosonic Branes with a Modified Measure}

The action of bosonic $p$-branes with a modified world-volume 
integration measure reads (cf. \ct{mstring-orig}):
\br
S_{p\mathrm{-brane}} = 
- \int d^{p+1}\s \,\P (\vp) \Bigl\lb \h \g^{ab} \pa_a X^{\m} \pa_b X^{\n} 
G_{\m\n}
\phantom{aaaaaaaaa} 
\nonu \\
+ \frac{\vareps^{a_1\ldots a_{p+1}}}{(p+1)\sqrt{-\g}}\( 
\pa_{a_1} X^{\m_1}\ldots \pa_{a_{p+1}} X^{\m_{p+1}} B_{\m_1\ldots\m_{p+1}}
- F_{a_1\ldots a_{p+1}} (A) \) \Bigr\rb \equiv - \int d^{p+1}\s \,\P (\vp) L
\lab{action-brane}
\er
\be
\P (\vp) \equiv \frac{1}{(p+1)!} \vareps_{i_1\ldots i_{p+1}} 
\vareps^{a_1\ldots a_{p+1}} \pa_{a_1} \vp^{i_1}\ldots \pa_{a_{p+1}} 
\vp^{i_{p+1}}
= \frac{1}{p!}\vareps_{ij_1\ldots j_p} \vareps^{\a_1\ldots \a_p}
{\dot{\vp}}^i \pa_{\a_1} \vp^{j_1} \ldots \pa_{\a_p} \vp^{j_p}
\lab{def-measure-brane}
\ee
Here the following notations are used:
\be
\underline{\s} \equiv \bigl( \s^a\bigr) \equiv \bigl(\s^0 \equiv \t,\s^\a\bigr)
\equiv \bigl(\t,\vec{\s}\bigr) \quad ; \quad
F_{a_1\ldots \a_{p+1}} (A) = (p+1)\pa_{\lb a_1} A_{a_2\ldots 
a_{p+1}\rb}
\lab{F-p-def}
\ee
where $a,b =0,1,\ldots,p$; $\a,\b =1,\ldots,p$;
$i,j =1,\ldots,p+1$; $\m,\n = 0,1,\ldots ,D-1$; $G_{\m\n}$ and
$B_{\m_1\ldots \m_{p+1}}$ denote space-time metric and antisymmetric 
$p+1$-rank tensor external fields, respectively. Also, it is convenient 
within the Hamiltonian formalism to introduce the following notations:
\be
\vareps^{\a_1\ldots \a_p} F_{0 \a_1\ldots \a_p} (A) =
\dot{\cA} - \pa_\a \cA^{\a}_0
\lab{tensor-strength} 
\ee
with:
\be
\cA \equiv \vareps^{\a_1\ldots \a_p} A_{\a_1\ldots \a_p} \quad ,\quad
\cA^{\a}_0 \equiv \vareps^{\a \b_1\ldots \b_{p-1}} A_{0 \b_1\ldots 
\b_{p-1}}
\lab{A-tensor-def}
\ee

In analogy with the string case we can put (closed) $(p-1)$-branes on the
world-volume of the modified-measure $p$-brane \rf{action-brane} 
coupled to the latter via the auxiliary world-volume $p$-form gauge field 
$A_{a_1\ldots a_p}$ giving rise to the following additional term in the
action \rf{action-brane} :
\br
S = S_{p\mathrm{-brane}} + S_{(p-1)\mathrm{-brane}} 
\nonu
\er
\br
S_{(p-1)\mathrm{-brane}} = \sum_i e_i
\int d^{p+1}\s\, A_{a_1\ldots a_p}(\underline{\s}) 
\int d^p u\, {1\o {p!}} \vareps^{\a_1\ldots\a_p}
\partder{\s_i^{a_1}}{u^{\a_1}}\ldots \partder{\s_i^{a_p}}{u^{\a_p}}
\d^{(p+1)} \bigl(\underline{\s} - \underline{\s}_i (\underline{u})\bigr)
\nonu \\
= \sum_i e_i \int d^{p+1}\s\, \cA^\a_0 (\underline{\s})
\int d^{p-1} u\, {1\o {(p-1)!}} \vareps_{\a\a_1\ldots\a_{p-1}}\,
{1\o {(p-1)!}}\vareps^{m_1\ldots m_{p-1}} 
\partder{\s_i^{\a_1}}{u^{m_1}}\ldots \partder{\s_i^{\a_{p-1}}}{u^{m_{p-
1}}}
\d^{(p)} \bigl(\vec{\s} - \vec{\s}_i (\vec{u})\bigr)
\lab{action-lower-brane}
\er
Here $\underline{u} \equiv \bigl( u^0 =\t,u^m\bigr) \equiv
\bigl(\t,\vec{u}\bigr)$ with $m=1,\ldots ,p-1$ are the world-volume
parameters of the pertinent $(p-1)$-branes embedded in the world-volume of the
original $p$-brane via the parameter equations
$\underline{\s} = \underline{\s}_i (\underline{u})$ (and we have chosen
the static gauge $\s^0 \equiv \t = u^0$ for all of them)\foot{In what follows
we assume that the $(p-1)$-branes do not intersect each other on the
original $p$-brane world-volume.}. Also, in the second equality 
\rf{action-lower-brane} we have used the notations from \rf{A-tensor-def}.

The Lagrangian formalism analysis of the modified-measure $p$-brane model
(without attached lower-dimensional branes) \rf{action-brane} has 
been performed in \ct{mstring-orig}. It parallels the analysis of the 
modified-measure string model (cf. Sect.1) where the analogues of 
Eqs.\rf{L-const}--\rf{g-eqs-motion} now read (taking for simplicity 
$B_{\m_1\ldots\m_{p+1}} = 0 $) :
\be
\h \g^{ab} \pa_a X^{\m} \pa_b X^{\n} G_{\m\n} 
- \frac{\vareps^{a_1\ldots a_{p+1}}}{(p+1)\sqrt{-\g}}
F_{a_1\ldots a_{p+1}} (A) = M
\lab{L-brane}
\ee
\be
\vareps^{a a_1\ldots a_p} \pa_a \Bigl(\frac{\P (\vp)}{\sqrt{-\g}}\Bigr) = 0
\quad \longrightarrow \quad 
\frac{\P (\vp)}{\sqrt{-\g}} = \mathrm{const} \equiv T
\lab{brane-tension}
\ee
\be
\pa_a X^{\m} \pa_b X^{\n} G_{\m\n} - \frac{\g_{ab}}{(p+1)\sqrt{-\g}}
\vareps^{a_1\ldots a_{p+1}} F_{a_1\ldots a_{p+1}} (A) = 0
\lab{g-eqs-motion-brane}
\ee
In Eq.\rf{L-brane} $M$ denotes arbitrary integration constant which enters
the relation between the intrinsic and the induced metrics on the $p$-brane
world-volume which follows from \rf{L-brane} and \rf{g-eqs-motion-brane}:
\be
\g_{ab} = \frac{p-1}{2M}\, \pa_a X^\m \pa_b X^\n G_{\m\n}
\lab{metrics-rel}
\ee
Also we have:
\be
\pa_a X^{\m} \pa_b X^{\n} G_{\m\n} - \frac{\g_{ab}}{p+1}\,
\g^{cd} \pa_c X^{\m} \pa_c X^{\n} G_{\m\n} = 0
\lab{Pol-like-eq-brane}
\ee

The arbitrariness of $M$ is due to the manifest invariance of the
modified-measure $p$-brane action \rf{action-brane} under the following
global scale symmetry \ct{mstring-orig} :
\be
\vp^i \to \l_i \vp^i \quad ,\quad 
\g_{ab} \to \bigl(\prod_i \l_i\bigr) \g_{ab} \quad ,\quad 
A_{a_1\ldots a_p} \to \bigl(\prod_i \l_i\bigr)^{{p-1}\o 2} A_{a_1\ldots a_p}
\lab{global-scale}
\ee
which can be used to fix the value of $M$, \textsl{e.g.}, $M =\h (p-1)$.
Note that the ``boundary'' term \rf{action-lower-brane} is not invariant under
the scale symmetry \rf{global-scale}, unless we simultaneously rescale the
``charge'' coupling constants $e_i$. Moreover, unlike the string case there
is no analogue of the $\P$-extended Weyl symmetry \rf{vp-diff}--\rf{gamma-conf}
for the modified-measure $p$-brane model \rf{action-brane}. The reason is due
to the fact that for $p\geq 2$ the standard measure density $\sqrt{-\g}$
transforms differently
~$\sqrt{-\g} \to \bigl( J(\vp)\bigr)^{{p+1}\o 2}\sqrt{-\g}$~ than the
modified measure density $\P (\vp)$ \rf{vp-diff} (cf. refs.\ct{G-3}).

The canonical Hamiltonian treatment of the $p$-brane model 
\rf{action-brane} with attached $(p-1)$-branes on its world-volume similarly 
follows the same steps as the canonical treament of the modified-measure string 
model in the previous Section. For the canonical momenta of 
$\vp^i,\, \cA,\, X^\m$ we have 
(using the short-hand notation $L$ from \rf{action-brane}) :
\be
\pi^{\vp}_i = - \vareps_{ij_1\ldots j_p} \vareps^{\a_1 \ldots \a_p}
\pa_{\a_1} \vp^{j_1}\ldots \pa_{\a_p} \vp^{j_p}\, L \quad ,\quad
\pi_{\cA} \equiv \cE = \frac{\P (\vp)}{\sqrt{-\g}}
\lab{momenta-brane-1}
\ee
\be
\cP_\m = \P (\vp) \Bigl\lb - 
\(\g^{00}\dot{X}^\n + \g^{0\a}\pa_\a X^\n\) G_{\m\n} - 
\frac{\vareps^{\a_1 \ldots \a_p}}{\sqrt{-\g}} 
\pa_{\a_1} X^{\n_1}\ldots \pa_{\a_p} X^{\n_p} B_{\m\n_1 \ldots \n_p} 
\Bigr\rb
\lab{momenta-brane-X}
\ee
Also, similarly to \rf{prim-constr} we have the following primary 
constraints:
\be
\pi_{\cA^\a_0} = 0 \quad ,\quad \pi_{\g^{ab}} = 0 \quad ,\quad 
\pa_\a \vp^i \pi^{\vp}_i = 0
\lab{prim-constr-brane}
\ee
where the last Virasoro-like constraints follow directly from the first
Eq.\rf{momenta-brane-1}.

At this point it is convenient to reexpress the world-volume Riemannian
metric $\g_{ab}$ in terms of its purely space-like part $\g_{\a\b}$
and the associated shift vector $N^\a$ and lapse function $N$ 
(see \textsl{e.g.} \ct{HRT}) :
\be
\g_{00} = - N^2 \bar{\g} + \bar{\g}_{\a\b} N^\a N^\b \quad ,\quad
\g_{0\a} = \bar{\g}_{\a\b} N^\b \quad ,\quad \bar{\g}_{\a\b} = \g_{\a\b}
\lab{lapse-shift}
\ee
where $\bar{\g} = \det\v\v\g_{\a\b}\v\v$. In particular,
$\sqrt{-\g} = N\bar{\g}$.

Using Eqs.\rf{momenta-brane-1}--\rf{momenta-brane-X} and the notations 
\rf{lapse-shift} we find the following canonical Hamiltonian 
(cf. Eq.\rf{Ham-canon}) :
% \br
% \cH = - \frac{1}{\sqrt{-\g} \g^{00}} \h \Bigl\lb
% \frac{G^{\m\n}}{\cE}\(\cP_\m + \cE \vareps^{\a_1\ldots \a_p}
% \pa_{\a_1} X^{\m_1} \ldots \pa_{\a_p} X^{\m_p} 
% B_{\m\m_1\ldots\m_p}\) \times
% \nonu \\
% \times \(\cP_\n + \cE \vareps^{\b_1\ldots \b_p}
% \pa_{\b_1} X^{\n_1} \ldots \pa_{\b_p} X^{\n_p} B_{\n\n_1\ldots\n_p}\) 
% + \cE \g \(\g^{00}\g^{\a\b} - \g^{0\a}\g^{0\b}\)
% G_{\m\n} \pa_\a X^\m \pa_\b X^\n \Bigr\rb
% \nonu \\
% + \frac{\g^{0\a}}{\g^{00}} \cP_\m \pa_\a X^\m 
% + \cE\sqrt{-\g}F(\pi^\vp ,\pa \vp) + \cE \pa_\a \cA^\a_0 +
% \Bigl(\; (p-1)-\mathrm{brane ~terms}\;\Bigr) 
% \lab{Ham-canon-brane}
% \er
\br
\cH = {N\o 2}\Bigl(\frac{G^{\m\n}}{\cE} \wti{\cP}_\m \wti{\cP}_\n +
\cE \bar{\g}\g^{\a\b} G_{\m\n} \pa_\a X^\m \pa_\b X^\n \Bigr) + 
\cE\, N\, F(\pi^\vp ,\pa \vp)
\nonu \\
- N^\a \wti{\cP}_\m \pa_\a X^\m +
\cE \pa_\a \cA^\a_0 +\Bigl(\; (p-1)-\mathrm{brane ~terms}\;\Bigr) 
\lab{Ham-canon-brane}
\er
with the short-hand notations:
\br
\wti{\cP}_\m \equiv \cP_\m + \cE \vareps^{\a_1\ldots \a_p}
\pa_{\a_1} X^{\m_1} \ldots \pa_{\a_p} X^{\m_p} B_{\m\m_1\ldots\m_p}
\nonu
\er
\be
F(\pi^\vp ,\pa \vp) \equiv (p-1)!
\sum_{i=2}^{p+1}\frac{\pi^{\vp}_i}{\vareps_{ij_1\ldots j_p}
\vareps^{\a_1\ldots \a_p} \pa_{\a_1} \vp^{j_1}\ldots \pa_{\a_p} 
\vp^{j_p}}
\lab{F-def}
\ee
and where the last terms in \rf{Ham-canon-brane} come from 
\rf{action-lower-brane}.

Commuting the canonical Hamiltonian \rf{Ham-canon-brane} with the primary
constraints \rf{prim-constr-brane} (where upon using the notations
\rf{lapse-shift} we have
$\pi_N\! =\! 0\, ,\,\pi_{N^\a}\! =\! 0\, ,\,\pi_{\bar{\g}^{\a\b}}\! =\! 0$
instead of $\pi_{\g^{ab}}\! =\! 0$) we obtain a set of secondary constraints.
Using the Poisson-bracket relation:
\be
\Pbr{\pa_\a \vp^i \pi^{\vp}_i (\vec{\s})}{F(\pi^\vp ,\pa \vp) (\vec{\s}^\pr)}
= - \d (\vec{\s}-\vec{\s}^\pr)\, \pa_a F(\pi^\vp ,\pa \vp) (\vec{\s})
\lab{Vir-1-brane}
\ee
we get the following secondary constraint:
\be
\pa_\a F(\pi^\vp ,\pa \vp) = 0  \quad \longrightarrow \quad
F(\pi^\vp ,\pa \vp) = - 2M \equiv \mathrm{const}
\lab{second-constr-brane}
\ee
where $M$ is arbitrary constant (it is the Hamiltonian counterpart of the
arbitrary integration constant $M$ appearing within the Lagrangian treatment,
cf. Eq.\rf{L-brane}). Once again, as in the string case, we find that the 
Virasoro-like constraints $\pa_\a \vp^i \pi^{\vp}_i$ together with
$F(\pi^\vp ,\pa \vp) + 2M =0$ (the latter being defined in \rf{F-def}) form a 
closed algebra of first class constraints implying that the auxiliary 
scalar fields $\vp^i$ are pure-gauge degrees of freedom.

Next, commuting \rf{Ham-canon-brane} with $\pi_{\cA^\a_0}$ yields:
\be
\pa_\a \cE (\underline{\s}) +
\sum_i e_i \int d^{p-1} u\, {1\o {(p-1)!}} \vareps_{\a\a_1\ldots\a_{p-1}}\,
{1\o {(p-1)!}}\vareps^{m_1\ldots m_{p-1}} 
\partder{\s_i^{\a_1}}{u^{m_1}}\ldots \partder{\s_i^{\a_{p-1}}}{u^{m_{p-1}}}
\d^{(p)} \bigl(\vec{\s} - \vec{\s}_i(\vec{u})\bigr) = 0
\lab{brane-Gauss-law}
\ee
which is the $p$-brane analog of the ``Gauss'' law constraint in the 
string case (second Eq.\rf{second-constr-a}). Further, since the canonical
Hamiltonian \rf{Ham-canon-brane} does not depend explicitly on $\cA$
\rf{A-tensor-def} (canonically conjugate to $\cE$), the $p$-brane ``electric''
field strength $\cE$ is conserved (world-volume time-independent) and as long as
it obeys the generalized ``Gauss law'' on the world-volume Eq.\rf{brane-Gauss-law},
$\cE$ is also world-volume {\em piece-wise} constant field with jumps along 
the normals equal to the ``charge'' $e_i$ when crossing the world-hypersurface
of the $i$-th $(p-1)$-brane. 

The rest of the secondary constraints reads:
\be
\frac{G^{\m\n}}{\cE} \wti{\cP}_\m \wti{\cP}_\n +
\cE \bar{\g}\g^{\a\b} G_{\m\n} \pa_\a X^\m \pa_\b X^\n - 4M\,\bar{\g}\,\cE
=0
\lab{brane-constr-A}
\ee
\be
\wti{\cP}_\m \pa_\a X^\m = 0 \quad ,\quad 
\pa_\a X^\m \pa_\b X^\n G_{\m\n} - \frac{2M}{p-1} \bar{\g}_{\a\b} = 0
\lab{brane-constr-B}
\ee
We now observe that for the special choice $M\! =\! \h (p-1)$, and provided
we identify the ``electric'' field strength $\cE$ as a dynamical brane tension
$T$, the constraints \rf{brane-constr-A}--\rf{brane-constr-B} coincide with the
secondary constraints within the Hamiltonian treatment of the usual 
Polyakov-like $p$-brane (the latter together with the primary constraints form a
mixture of first-class and second-class constraints).

Thus, we conclude that the modified-measure $p$-brane model
\rf{action-brane} possesses, apart from the same brane degrees of freedom as
the standard Polyakov-like $p$-brane, an additional brane degree of freedom
$\cE$ -- an world-volume ``electric'' field strength, which can be identified 
as a dynamical brane tension and which, according to Eq.\rf{brane-Gauss-law},
may be variable in general.
\lskip
{\large{\textbf{4. Superstrings with a Modified Measure}}}
% \section{Superstrings with a Modified Measure}

We consider the following Green-Schwarz-type of superstring action 
with a modified world-sheet integration measure:
\be
S_{\mathrm{superstring}} = 
\int d^2\s \,\P (\vp) \Bigl\lb - \h \g^{ab} \Pi_a^{\m} \Pi_{b\,{\m}} +
\frac{\vareps^{ab}}{\sqrt{-\g}}\Bigl( \Pi_a^{\m} \bigl(\th\s_\m 
\pa_b\th\bigr) + \h F_{ab} (A) \Bigr) \Bigr\rb \equiv - \int d^2\s \,\P (\vp) L
\lab{action-sstring}
\ee
with the same notations as in \rf{action-string} and \rf{def-measure} (for
simplicity we take now $G_{\m\n}=\eta_{\m\n},\, B_{\m\n}=0$) and where:
\be
\Pi_a^{\m} \equiv \pa_a X^{\m} + i \th \s^\m \pa_a\th
\lab{Pi-def}
\ee
Here $\th \equiv (\th^\a )$ ($\a = 1,\ldots ,16$) denotes 16-dimensional
Majorana-Weyl spinor in the embedding $D\! =\! 10$ space-time, whereas
$\s^\m \equiv \( (\s^\m)_{\a\b}\)$ indicate the upper diagonal $16\times 16$
blocks of the $32\times 32$ matrices $\cC^{-1}\G^\m$ with $\G^\m$ and
$\cC$ beying the $D\! =\! 10$ Dirac and charge-conjugation matrices,
respectively.

The Lagrangian in \rf{action-sstring} is explicitly invariant under
space-time supersymmetry transformations:
\be
\d_\eps \th = \eps \quad,\quad \d_\eps X^{\m}= -i \bigl(\eps \s^{\m} 
\th\bigr)
\quad ,\quad 
\d_\eps A_a = i \bigl(\eps \s_{\m} \th\bigr) \(\pa_a X^{\m} +
{i\o 3} \th \s^{\m} \pa_a\th\)
\lab{susy-transf}
\ee
In particular, the algebra of supersymmetry transformations 
\rf{susy-transf} closes on $A_a$ up to a gauge transformation:
\be
\lcurl \d_{\eps_1},\, \d_{\eps_2}\rcurl A_a =
\pa_a \Bigl( -{2\o 3}\bigl(\eps_1\s^\m \th\bigr) 
\bigl(\eps_2\s_\m \th\bigr)\Bigr)
\lab{susy-close}
\ee

Let us note that the action \rf{action-sstring} bears resemblance to the
modified Green-Schwarz superstring action proposed by Siegel \ct{Siegel-GS}
provided we replace the modified integration measure density $\P (\vp)$ with
the ordinary one $\sqrt{-\g}$ and provided we redefine the auxiliary gauge
field $A_a$ as fermionic bilinear composite $A_a = - i \th_\a \pa_a \phi^\a$
(cf. second ref.\ct{mstring-orig})
with $\phi$ indicating Siegel's auxiliary fermionic world-sheet field
which is a space-time spinor similar to $\th$. However, let us emphasize
that our present approach to the modified-measure superstring model 
\rf{action-sstring} is consistently based on a fundamental (non-composite) gauge
field $A_a$.

For the canonical momenta of $\vp^i,\, A_1,\, X^\m, \th$ we have (using the
short-hand notation $L$ from \rf{action-sstring} and \rf{Pi-def}) :
\be
\pi^{\vp}_i = - \vareps_{ij} \pa_\s \vp^j L \quad ,\quad
\pi_{A_1} \equiv E = \frac{\P (\vp)}{\sqrt{-\g}}
\lab{smomenta-1}
\ee
\be
\cP_\m = \P (\vp) \Bigl\lb - 
\(\g^{00} \Pi_{0\,\m} - \g^{01} \Pi_{1\,\m}\) +
\frac{i}{\sqrt{-\g}} \bigl(\th \s_\m \th^{\pr}\bigr)\Bigr\rb
\lab{smomenta-X}
\ee
\be
\cP_\th = \P (\vp) \Bigl\lb - 
\(\g^{00} \Pi_{0\,\m} - \g^{01} \Pi_{1\,\m}\) -
\frac{i}{\sqrt{-\g}} X_\m^\pr\Bigr\rb\, i \th \s^\m
\lab{smomenta-th}
\ee
where the prime now indicates the derivative $\pa_\s$. From 
\rf{smomenta-X}--\rf{smomenta-th} and taking into account the second
Eq.\rf{smomenta-1} we obtain the fermionic primary constraint:
\be
i\cD \equiv \cP_\th - \Bigl(\cP_\m - E \Pi_{1,\,\m}\Bigl)\, i \th \s^\m = 0
\lab{kappa-constr}
\ee
Therefore, we have the following set of primary constraints:
\be
\pi_{A_0} = 0 \quad ,\quad \pi_{\g^{ab}} = 0 \quad ,\quad 
\pa_\s \vp^i \pi^{\vp}_i = 0 \quad ,\quad \cD = 0
\lab{sprim-constr}
\ee
Now, for the velocities as functions of the canonical coordinate and 
momenta we get:
\be
\dot{X}^\m + i \th\s^\m \dot{\th} \equiv \Pi^\m_0 (\ldots) =
\frac{1}{\sqrt{-\g}\g^{00}} \Bigl(\frac{-\cP^\m}{E} + i\th\s^\m 
\th^{\pr}\Bigr)
- \frac{\g^{01}}{\g^{00}} \Pi^\m_1
\lab{sX-dot}
\ee
\br
\dot{A}_1 - i \Pi^\m_1 \bigl(\th \s_\m \dot{\th}\bigr) 
\equiv \dot{\cA}_1 (\ldots )
= \pa_\s A_0 - \sqrt{-\g} \frac{\pi^{\vp}_2}{\pa_\s \vp_1} +
\nonu \\
\sqrt{-\g} \Bigl(\h\g^{00} \Pi_0^\m (\ldots) \Pi_{0\,\m} (\ldots) 
+ \g^{01} \Pi_0^\m (\ldots) \Pi_{1\, m} + \h \g^{11} \Pi^\m_1 \Pi_{1\,m}\Bigr)
- i \bigl(\th\s_\m\th^\pr\bigr) \Pi_0^\m (\ldots)
\lab{sA1-dot}
\er
In Eq.\rf{sA1-dot} we used the short-hand notation $\Pi^\m_0 (\ldots )$
defined in \rf{sX-dot}. The canonical Hamiltonian reads:
\be
\cH = \cP_\m \dot{X}^\m (\ldots) + \cP_\th \dot{\th} (\ldots) 
+ E \dot{A}_1 (\ldots) + i\L_\a \cD^\a =
\cP_\m \Pi^\m_0 (\ldots) + E \dot{\cA}_1 (\ldots) + 
i\cD \(\dot{\th}(\ldots) - \L\)
% + {\mathrm{boundary terms}}
\lab{sHam-canon-0}
\ee
Here $(\ldots)$ indicate that all velocities as considered as functions 
of the canonical coordinate and momenta according to 
\rf{sX-dot}--\rf{sA1-dot}; $\cD$ is the fermionic primary constraint 
\rf{kappa-constr} and $\L$ is the corresponding fermionic Lagrange multiplier
which is determined from the requirement of the preservation of the constraint
$\cD$ under the Hamiltonian dynamics by \rf{sHam-canon-0}. Inserting in 
\rf{sHam-canon-0} the expressions \rf{sX-dot}--\rf{sA1-dot} we obtain:
\br
\cH = -\frac{1}{\sqrt{-\g}\g^{00}} \h \Bigl\lb \frac{1}{E}
\(\cP^\m - iE\bigl(\th\s^\m\th^\pr\bigr)\)
\(\cP_\m - iE\bigl(\th\s_\m\th^\pr\bigr)\) + E \Pi^\m_1 
\Pi_{1\,\m}\Bigr\rb
\nonu \\
+ \frac{\g^{01}}{\g^{00}} \(\cP_\m - iE\bigl(\th\s_\m\th^\pr\bigr)\) \Pi^\m_1
+i\L\cD + E \pa_\s A_0 - E \sqrt{-\g} \frac{\pi^{\vp}_2}{\pa_\s \vp_1} 
% + {\mathrm{boundary terms}}
\lab{sHam-canon}
\er
Commuting of the canonical Hamiltonian \rf{sHam-canon} with the primary 
constraints \rf{sprim-constr} leads to the following secondary constraints:
\be
\frac{\pi^{\vp}_2}{\pa_\s \vp_1} = 0 \quad ,\quad
\pa_\s E =0  \quad \Bigl(\; {\textrm{``Gauss ~law''}}\; \Bigr) 
\lab{s-second-constr-a}
\ee
\be
\cT_{+} \equiv {1\o 4}\Bigl\lb \frac{\cP}{E} + 
E\bigl(X^\pr - 2i\th\s\th^\pr\bigr)\Bigr\rb^2 - i\th^\pr \cD = 0
\quad ,\quad
\cT_{+} \equiv {1\o 4}\Bigl(\frac{\cP}{E} - E X^\pr\Bigr)^2
\lab{s-second-constr}
\ee
Therefore, as in the purely bosonic case we conclude that the 
canonical Hamiltonian is a linear combination of constraints only.

As in the bosonic case, the constraints involving the auxiliary scalar 
fields $\vp^i$ span the same Poisson-bracket algebra \rf{Vir-alg}--\rf{Vir-1} 
and, therefore, the auxiliary scalars are again pure-gauge degrees of 
freedom. The rest of the constraint algebra is the same as in the case of the 
standard Green-Schwarz formulation\foot{Let us recall that the fermionic spinor
constraint $\cD$ \rf{kappa-constr} contains a Lorentz non-covariant 
mixture of first-class (``kappa''-symmetry) and second-class constraints. To 
solve the problem of super-Poincare covariant quantization of the standard
Green-Schwarz superstring a new reformulation of the latter has been
proposed in refs.\ct{NPS} involving a special set of auxiliary bosonic
pure-gauge world-sheet scalar fields (``harmonic'' variables). For 
recent developments on this subject, see \ct{GS-cov-recent} and references 
therein.}
provided (in full analogy with the purely bosonic case) we identify the 
world-sheet ``electric'' field strength $E$ as dynamically generated string 
tension $T$.
\lskip
{\large{\textbf{5. Strings with ``$\P$-Extended Weyl Invariant'' Action 
for Non-Abelian World-Sheet Gauge Field}}}
% \section{Strings with the "$\Phi$ extended conformal invariant" 
% action for non-Abelian World-Sheet Gauge Field}
\lskip

% \subsection{The regular measure version of the theory}
\textbf{5.1 The Regular-Measure Version of the Theory}

As it is well known, in four space-time dimensions the standard gauge 
field action
$\propto\int\sqrt{-g} d^{4}x Tr(F_{\mu\nu}F^{\mu\nu})$ is invariant 
under transformations $g_{\mu\nu}\rightarrow \Omega^{2}(x)g_{\mu\nu}$, 
\textsl{i.e.}, it is conformally invariant. In $D=2$, the appropriate 
conformally invariant action, provided we use the standard measure 
$\sqrt{-\gamma}$, would be:
\be
\int d^2\s\, \sqrt{-\g} \sqrt{-\h \Tr (F_{ab}(A) F_{cd}(A)) \g^{ac}\g^{bd}}
= \int d^2\s\, \sqrt{\Tr (F_{01}(A) F_{01}(A))}   
\lab{BI-top-action}
\ee
where: 
\be
F_{ab}(A) = \pa_a A_b - \pa_b A_c + i \bigl\lb A_a,\, A_b\bigr\rb
\lab{F-NA-def}
\ee
is a non-Abelian world-sheet gauge field-strength and we have used 
$F_{ab}(A) = \vareps_{ab} F_{01}(A)$. As we see, the action \rf{BI-top-action}
is not only independent of the conformal factor in the metric, but also it is
totally metric independent, \textsl{i.e.}, the $D\! =\! 2$ 
``square-root Yang-Mills'' model \rf{BI-top-action} is topological in the same
sense as, \textsl{e.g.}, the $D\! =\! 3$ Chern-Simmons model. Due to this fact
the string and gauge degrees of freedom turn out to be decoupled.

To see that such theory does not lead to a well defined dynamics and
instead a modified-measure version of \rf{BI-top-action} is necessary,
we consider first the equations of motion that result from
\rf{BI-top-action}. Variation with respect to gauge fields $A_{a}$
yields:
\be
\nabla_a \Bigl(\frac{F_{01}}{\sqrt{\Tr (F_{01} F_{01})}}\Bigr) = 0
\lab{A-eqs-NA-1}
\ee
or, equivalently:
\be
\nabla_a F_{01} - F_{01} \frac{\Tr (F_{01} \nabla_a F_{01})}{\Tr 
(F_{01}
F_{01}} =0
\lab{A-eqs-NA-2}
\ee
which in turn are equivalent to the equations:
\be
\nabla_a F_{01} = \pa_a f \, F_{01}
\lab{A-eqs-NA-3}
\ee
with $f\equiv f(\t,\s)$ being an arbitrary colorless world-sheet scalar
field. The general solution of \rf{A-eqs-NA-3} reads:
\be
F_{01} = G^{-1} e^{f(\t,\s)}\cM_0 G
\lab{BI-F-sol}
\ee
\be
A_0 = G^{-1} \Big( -\cM_0 \int^\s \!\!\! d\s^\pr e^{f(\t,\s^\pr)} \Bigr) G
-i G^{-1}\pa_\t G      \quad ,\quad
A_1 = -i G^{-1}\pa_\s G
\lab{BI-A-sol}
\ee
where $G$ is arbitrary $(\t,\s)$-dependent element of the gauge group
(reflecting the gauge freedom) whereas $\cM_0$ is arbitrary constant
element of the corresponding Lie algebra.
   
Thus, we see that in the $D\! =\! 2$ ``square-root Yang-Mills'' action
\rf{BI-top-action} there is an additional freedom in equations of motion 
(beyond the usual non-Abelian gauge symmetry) which is manifested in the
appearance of the arbitrary (not determined by the dynamics) world-sheet 
scalar field $f(\t,\s)$ in \rf{A-eqs-NA-3}--\rf{BI-A-sol}. 

This can be equivalently understood from the canonical Hamiltonian 
point of view. Namely, one can show that the canonical Hamiltonian of the 
$D\! =\! 2$ ``square-root Yang-Mills'' model \rf{BI-top-action} is a 
linear combination of first-class constraints only in contrast to the ordinary 
Yang-Mills case:
\be
\cH = \Tr\Bigl(\cE \(\pa_\s A_0 + i\Sbr{A_1}{A_0}\)\Bigr) + \L_0\,
\pi_{A_0} +
{\L\o 2}\(\Tr \cE^2 - 1\)
\lab{BI-Ham}  
\ee
where $\pi_{A_0}$ and $\cE = \frac{F_{01}}{\sqrt{\Tr (F_{01} F_{01})}}$
are the canonical momenta of $A_0$ and $A_1$, respectively, and where
$\L_0,\, \L$ are the corresponding Lagrange multipliers. Notice the
appearance of the third first-class constraint term in \rf{BI-Ham} 
instead of the standard non-constraint term $\h \Tr \cE^2$. Moreover, the 
total number of first-class constraints in \rf{BI-Ham} exceeds the number of 
the underlying degrees of freedom.
\lskip 
% \subsection{Modified Measure Version (the case of closed strings
% without charges)}

\textbf{5.2 Modified-Measure Version -- The Case of Closed Strings
without Charges}

We will now see that the modified-measure version of Non-Abelian
world-sheet gauge fields has a well defined dynamics (in contrast to the
regular measure case of the previous subsection) provided that the theory
possesses the $\P$-extended Weyl symmetry.
We consider the following  non-Abelian generalization of the original
bosonic string action with a modified measure \rf{action-string} (now 
we take for simplicity $G_{\m\n} = \eta_{\m\n}$ and $B_{\m\n} = 0$) :
\br
S = - \int d^2\s \,\P (\vp) \Bigl\lb \h \g^{ab} \pa_a X^{\m} \pa_b X_\m -
\sqrt{-\h \Tr (F_{ab}(A)F_{cd}(A)) \g^{ac}\g^{bd}}\Bigr\rb
\nonu
\er
\be
= - \int d^2\s \,\P (\vp) \Bigl\lb \h \g^{ab} \pa_a X^{\m} \pa_b X_\m -
{1\o \sqrt{-\g}}
\sqrt{\Tr (F_{01}(A)F_{01}(A))}\Bigr\rb
\equiv - \int d^2\s \,\P (\vp) L
\lab{action-string-NA}
\ee
where $F_{ab}(A)$ is the non-Abelian world-sheet gauge field-strength as in
\rf{F-NA-def}.

Similar to what we have seen in Sec.1, the variation with respect to 
the measure $\P$ degrees of freedom $\vp^{i}$ leads to the equation
(provided that $\P\neq 0$):
\be
\h \g^{ab}\pa_a X^\m \pa_b X_\m -\frac{1}{\sqrt{-\g}}
\sqrt{\Tr (F_{01} F_{01})} = M
\lab{vp-eqs}
\ee
Varying the action \rf{action-string-NA} with respect to $\gamma^{ab}$
we get: 
\be
\pa_a X^\m \pa_b X_\m -\frac{1}{\sqrt{-\g}}\g_{ab}\sqrt{\Tr (F_{01} F_{01})}=0
\lab{vargab-eqs}
\ee
Contracting this equation with $\g^{ab}$ and comparing with \rf{vp-eqs}
we conclude that again, similar to what has been shown in the simpler 
model of Section 1, $M=0$ and we obtain finally:
\be
\h \sqrt{-\g} \g^{ab}\pa_a X^\m \pa_b X_\m = \sqrt{\Tr (F_{01} F_{01})} 
\lab{vp0-eqs}
\ee
\be
\pa_a X^\m \pa_b X_\m - \g_{ab}\h \g^{cd}\pa_c X^\m \pa_c X_\m = 0
\lab{g-eqs-0}
\ee

Varying the action \rf{action-string-NA} with respect to $A_a$ we 
obtain:
\be
\nabla_a \cE \equiv \pa_a \cE + i \bigl\lb A_a,\,\cE \bigr\rb = 0 \quad ,
\quad
\cE \equiv \frac{\P (\vp)}{\sqrt{-\g}}
\frac{F_{01}}{\sqrt{\Tr (F_{01} F_{01})}}
\lab{A-eqs-NA}
\ee
with $\cE$ being the non-Abelian electric field-strength -- the 
canonically conjugated momentum of $A_1$. Accordingly,
Eq.\rf{A-eqs-NA} for $a=1$ represents the non-Abelian Gauss law on the
world-sheet. Using Eqs.\rf{A-eqs-NA} one can easily show:
\be
0 = \Tr \Bigl(\cE \nabla_a \cE\Bigr) = \h \pa_a \( \Tr \cE^2\)
=\h \pa_a \Bigl(\frac{\P (\vp)}{\sqrt{-\g}}\Bigr)^{2}
\lab{cE-sol}
\ee
\textsl{i.e.}, the ratio of the measure densities (the magnitude of the 
non-Abelian electric field-strength), which plays the role of a dynamically
generated string tension, is again constant:
$|\cE | \equiv | \frac{\P (\vp)}{\sqrt{-\g}}| = \mathrm{const}$.
The equations of motion \rf{A-eqs-NA} , upon using this fact,
coincide with the equations \rf{A-eqs-NA-2} (Eqs.\rf{A-eqs-NA-3}--\rf{BI-A-sol} 
similarly hold). However, in contrast to the regular measure version of 
the theory, now in the context of the modified-measure model 
\rf{action-string-NA} we have the equation \rf{vp0-eqs} which upon substituting
the solution \rf{BI-F-sol} in
$\sqrt{\Tr (F_{01} F_{01})} = e^{f(\t,\s)} \sqrt{\Tr\cM_0^2}$ determines
completely the function $f(\t,\s)$ in terms of the string solution.
\lskip
% \subsection{Charges, strings and classical mechanism for confinement}

\textbf{5.3 Charges, Strings and Classical Mechanism for 
Confinement}

Classical treatment of strings in the context of the Polyakov approach
(with the regular measure density $\sqrt{-\g}$) allows two possibilities for 
the string topology: the first one is a closed string where the string tension
is a constant all over the string; the second possibility is an open
string with end-points (and/or \textsl{ad hoc} with point-like charges at 
the end-points).

In the modified-measure string theory there are more possibilities due 
to the dynamical mechanism of tension generation. In fact, for both cases, 
\textsl{i.e.}, for closed and open strings, one can study models where one or
more point-like charges $C_{i}$, in general {\em non-Abelian} ``color'' ones,
are located inside the string\foot{Generically
one can consider smooth charge or current distributions along the string. 
Such more general cases we will study elsewhere; see also Apendix.}.
A simple model describing this situation consists of adding to the 
action \rf{action-string-NA} the following interaction term:
\be
S_{int}=-\sum_{i}\int\frac{d\s^{a}}{d\t_{i}}\Tr(C_{i}A_{a})d\t_{i}
\lab{inter}
\ee
where $\t_i$ indicate the corresponding proper times.
In the simplest case of {\em static} ``color'' charges $C_{i}$ localized 
at the points $\s_{i},\; (i=1,2,...)$, Eq.\rf{inter} reads:
\be
S_{int,static}=-\sum_{i}\Tr C_{i}\int d\t A_{0}(\t ,\s_{i})
\lab{inter,static}
\ee
The only changes in the equations of motion, comparing to the equations of the
previous subsection, occur in Eq.\rf{A-eqs-NA} which in the axial gauge 
($A_{1}=0$) take the form:
\be
\pa_\s \cE -\sum_{i} C_{i} \d (\s - \s_{i}) = 0
\lab{dglaw-nonabelian}
\ee
with $\cE$ as defined in the second Eq.\rf{A-eqs-NA}.

Let us first consider the solution of the ``Gauss law'' 
Eq.\rf{dglaw-nonabelian} in the case with two static point-like (color) 
charges $C_{1}$ and $C_{2}$ localized at the points $\s_{1}$ and $\s_{2}$
with $\s_{1} < \s_{2}$. To get this solution we perform integration in
\rf{dglaw-nonabelian} over $\s$ from some $\s < \s_{1}$ up to some 
$\s > \s_{2}$. Then we obtain:
\be
\cE (\s) = \left\{ \begin{array}{lll}
          \cE_{1} & \mbox{ for $\s<\s_{1}$}\\
          \cE_{2} & \mbox{ for $\s_{1}<\s<\s_{2}$}\\
          \cE_{3} & \mbox{ for $\s>\s_{2}$}  \end{array}  \right.
\qquad \mathrm{and} \qquad 
\cE_{2}-\cE_{1} = C_{1} \quad ,\quad \cE_{3}-\cE_{2} = C_{2} 
\lab{disc-boson}
\ee

To realize the physical case of such an open string (no periodic boundary 
conditions in $\s$ are assumed) with finite energy we have to consider a
finite string which is possible only if $\cE_{1}\equiv\cE_{3}\equiv 0$.
Then the charges $C_{1}$ and $C_{2}$ appears to be the end-points and it
follows from \rf{disc-boson} that:
\be
C_{1}+C_{2}=0 \quad \mathrm{and} \quad \cE_{2}=C_{1}
\lab{0color-boson}
\ee
Therefore, Eq.\rf{0color-boson} becomes the statement for color 
confinement of the two point-like charges $C_{i}$ (``quarks'') in a colorless
``meson-like state'' as a result of the variable dynamical tension of the 
string connecting them.

In a similar way one can construct a classical string model for baryons.
Let us consider a {\em closed} string parametrized by 
$\s\,\;(0\leq\s\leq 2\pi)$ with three static point-like color charges
$C_{1}$, $C_{2}$, $C_{3}$ localized at 
the points $\s_{1}$, $\s_{2}$, $\s_{3}$, respectively. Then solving 
Eq.\rf{dglaw-nonabelian} we obtain for the ``chromoelectric'' field,
\textsl{i.e.}, the dynamical string tension \rf{A-eqs-NA} :
\be
\cE (\s) = \left\{ \begin{array}{lll}
          \cE_{12} & \mbox{ for $\s_{1}=0<\s<\s_{2}$}\\
          \cE_{23} & \mbox{ for $\s_{2}<\s<\s_{3}$}\\
          \cE_{31} & \mbox{ for $\s_{3}<\s<2\pi$}
\end{array} \right.
\lab{E-baryon}
\ee
where $\cE_{12}$, $\cE_{23}$, $\cE_{31}$ are constants, which 
implies:    
\be
\cE_{12}-\cE_{31} = C_{1} \quad ,\quad \cE_{23}-\cE_{12} = C_{2} 
\quad ,\quad  \cE_{31}-\cE_{23} = C_{3}
\lab{disc-baryon}
\ee
Summing Eqs.\rf{disc-baryon} we get:
\be
C_{1}+C_{2}+C_{3}=0 
\lab{0color-baryon}
\ee  
which means that color confinement appears again, now in the case 
of a ``baryon-like'' configuration.

Notice that not only the orientations of $\cE_{12}$, $\cE_{23}$, $\cE_{31}$
in color space, but also their magnitude are different in general. The last
statement follows from the fact that Eq.\rf{cE-sol} does not hold in the
points where the charges are located. This means that the charges can be
sources of discontinuities of the tension (notice that the second equation
in \rf{A-eqs-NA} still holds).  This is possible here precisely due to the
identification of the string tension with the ratio of the measure
densities $\frac{\P (\vp)}{\sqrt{-\g}}$ (second Eq.\rf{A-eqs-NA})
being also the magnitude of the pertinent world-sheet ``chromoelectric''
field strength.
Due to these properties we may call the above modified-measure string model
with a $\P$-extended Weyl-invariant non-Abelian world-sheet gauge field action
\rf{action-string-NA} a {\em ``color'' string model}.

The above simple picture of point-like charge confinement via ``color'' strings
can be straightforwardly generalized to the case of higher-dimensional branes.
Namely, let us consider $N$ non-intersecting ``charged'' closed $(p-1)$-branes
living on a closed $p$-brane whose dynamics is governed by the
modified-measure brane action \rf{action-brane} and \rf{action-lower-brane}.
Let us also recall that the dynamically generated brane tension $\cE$ (cf.
second Eq.\rf{momenta-brane-1}) obeys the brane ``Gauss law'' constraint
Eq.\rf{brane-Gauss-law}. Denoting by $\cE_i$ the constant value of $\cE$ in
the strip on the fixed-time world-hypersurface of the $p$-brane situated
between the $(i-1)$-th and the $i$-th ``charged'' $(p-1)$-branes we find from
\rf{brane-Gauss-law} :
\be
\cE_{i+1} = \cE_i + e_i  \quad ,\quad i=0,1,\ldots ,N \quad 
\mathrm{with} \quad \cE_0 \equiv \cE_N \;\; ,\;\; e_0 \equiv e_N
\lab{E-brane-hadron}
\ee
Summing up Eqs.\rf{E-brane-hadron} we find similarly to the string case that
the only possible configuration of static ``charged'' closed $(p-1)$-branes
coupled pair-wise via modified-measure $p$-branes \rf{action-brane} is the
zero-charge one.
\lskip
{\large{\textbf{6. Discussion and Conclusions}}}

We have seen above how modifying the world-sheet (world-volume) 
measure of integration can significantly affect the implications of string
and brane dynamics. First of all, it turns out that to get an acceptable 
dynamics, the corresponding string and brane theories need the introduction of 
auxiliary world-sheet gauge field (world-volume $p$-form tensor gauge field). 
Furthermore, the tension of the string or brane is not any more a fundamental 
parameter (\textsl{i.e.}, a given \textsl{ad hoc} scale): it is dynamically 
determined as the magnitude of the pertinent gauge field strength and it is 
proportional to the ratio of the measure densities $\P/\sqrt{-\g}$. 
If no charges exist on the world-sheet (world-volume) then for closed 
strings (branes) the standard Polyakov type equations are obtained and the
Poisson-bracket algebra of the relevant Hailtonian constraints is
the same as that of the standard string (brane) theory. The same result holds 
also for the modified-measure superstring model.  

The string tension is identified as the canonically conjugate momentum of
the spatial component of the auxiliary world-sheet gauge potential,
therefore, it assumes the role of an ``electric'' field strength.
The latter is shown to obey the ``Gauss law'' equation. Thus, in the 
presence of world-sheet charges, the string tension can change dynamically. 
The latter becomes possible since the tension,\textsl{i.e.}, the ``electric''
field strength is proportional to the ratio of the measure densities 
$\P /\sqrt{-\g}$. In particular, point-like charges living on the string can 
be responsible for discontinuous changes of the string tension. The special 
case, when the string tension changes from a finite value to zero, can be 
regarded as the formation  of an ``edge'' on the string or, equivalently, as a
new way of formulating open strings. We have shown that similar results 
hold also for modified-measure theories of $p$-branes. Namely, $p$-form 
(tensor gauge) charges living on the $p$-brane, in particular, 
lower-dimensional ``charged'' $(p-1)$-branes lead to a dynamically changing 
brane tension.

Finally, we studied a conformally (Weyl-) invariant modified-measure string 
theory with non-Abelian gauge (``square-root Yang-Mills'') field living on the 
string world-sheet called ``color'' string. As a result, a simple classical 
mechanism for ``color'' confinement of point-like ``color'' charges via 
``color'' strings is proposed with the colorlessness of the corresponding
composite ``hadrons'' automatically emerging due to the new dynamics
inherent in the modified-measure string model. Similar picture of
confinement and colorlessness arises also for systems of ``charged'' 
$(p-1)$-branes coupled via modified-measure $p$-branes.

As a byproduct, it is found that a nice geometrical meaning can be given for the
auxiliary string world-sheet gauge fields: if these are of the abelian 
type, they can represent the world-sheet spin-connection associated with the 
(Abelian in $1+1$ dimensions) Lorentz group (see Eq.\rf{spin-curv} above). 

Notice that world-sheet gauge fields have also been considered in the very
interesting work \ct{Bergshoeff-etal}. In the latter case, however, a Nambu-Goto 
approach is employed so that the issue of conformal invariance peculiar to 
the Polyakov formulation is lost.
\lskip
\textbf{Acknowledgements.} 
We want to thank Zvi Bern for useful discussions and encouragement.
E.I.G. wants to thank the Physics Department at UCLA for its hospitality during
the last stages of the research presented here. E.N. and S.P. are partially 
supported by Bulgarian NSF grant {\sl F-904/99} and by a travel grant from 
Bulgarian Academy of Sciences.
\lskip
{\large{\textbf{Appendix. Strings with a Modified Measure Coupled to 
World-Sheet Currents}}}

Let us briefly discuss the case of bosonic strings with a modified
world-sheet integration measure coupled to an external space-time 
dilaton field. The pertinent action reads:
\be
S = - \int d^2\s \,\P (\vp) \Bigl\lb 
\h \g^{ab} \pa_a X^{\m} \pa_b X^{\n} G_{\m\n}(X) -
R(\om) \cU (X) \Bigr\rb \equiv - \int d^2\s \,\P (\vp) L
\lab{action-string-Dilaton}
\ee
with $R(\om)$ being the scalar curvature of the $D\! =\! 2$ spin 
connection $\om_a$ defined in Eq.\rf{spin-curv}.
Varying \rf{action-string-Dilaton} with respect to $\om_a$ we obtain 
once again dynamically generated string tension as:
\be
E \equiv \pi_{\om_1} = \frac{\P (\vp)}{\sqrt{-\g}} \cU (X) = 
\mathrm{const} \equiv T
\lab{momenta-1-Dilaton}
\ee
with $\pi_{\om_1}$ being the canonically conjugated momentum of $\om_1$,
which brings the action \rf{action-string-Dilaton} to the form:
\be
S = - T \int d^2\s \,\h \sqrt{-\g} \g^{ab} \pa_a X^\m \pa_b X^\n 
\frac{G_{\m\n}(X)}{\cU (X)}
\lab{action-string-Dilaton-1}
\ee
\textsl{i.e.}, an action describing string motion in a conformally 
modified extenal space-time background with 
$G^\pr_{\m\n}(X) = G_{\m\n}(X)/\cU (X)$. Thus, the model
\rf{action-string-Dilaton} differs significantly from the ordinary
Polyakov-type string coupled to a dilaton:
\be
S = - T \int d^2\s \,\h \sqrt{-\g} \Bigl\lb \g^{ab} \pa_a X^\m \pa_b X_\m 
+ R(\g) \cU (X) \Bigr\rb
\lab{action-string-Dilaton-Polyakov}
\ee

Now, let us consider a generalization of the string model \rf{action-string}
describing the coupling of the bosonic modified-measure string
through the auxiliary gauge field $A_a$ to a conserved world-sheet current
$\vareps^{ab} \pa_b v$ where $v$ is a world-sheet scalar field:
\be
S = - \int d^2\s \,\P (\vp) \Bigl\lb \h \g^{ab} \pa_a X^{\m} \pa_b X_{\m} +
\h \g^{ab} \pa_a v \pa_b v -
\frac{\vareps^{ab}}{2\sqrt{-\g}} F_{ab} (A) \Bigr\rb
+ \eps \int d^2\s \, A_a \vareps^{ab} \pa_b v
\lab{action-string-v}
\ee
Notice that the last term in Eq.\rf{action-string-v} can be rewritten
in the form:
\be
\eps \int\!\!\sqrt{-\g} d^2\s\, A_a \frac{\vareps^{ab}}{\sqrt{-\g}} \pa_b v
\lab{action-string-v-last}
\ee
which means that by including this term we study a model which belongs to the
class of {\em Two Measures Theories} \ct{GK-1} (TMT)\foot{In $D$-dimensional 
space-time the action has generically the form:
\be
S=\int d^{D}x \,\P (\vp) L_{1} +\int d^{D}x \,\sqrt{-\g} L_{2}
\lab{TMT}
\ee
where the Lagrangian densities $L_{1}$ and $L_{2}$ are independent of the
degrees of freedom $\vp^i$ building up $\P$.}. 

The equations of motion with respect to $A_a$ read:
\be  
\vareps^{ab} \pa_b \Bigl(\frac{\P (\vp)}{\sqrt{-\g}} + \eps v\Bigr) = 0
\quad ,\quad i.e. \;\; \frac{\P (\vp)}{\sqrt{-\g}} = C - \eps v
\lab{A-eqs}
\ee
where $C$ is a dynamically generated constant scale.
The canonical Hamiltonian treatment of \rf{action-string-v} is completely
analogous to the simpler case of \rf{action-string} in Sect.2. In
particular, for the auxiliary ``electric'' field strength we obtain:
\be
\pi_{A_1} \equiv E = \frac{\P (\vp)}{\sqrt{-\g}} \qquad \longrightarrow
\qquad
E + \eps v = C
\lab{momenta-1-v}
\ee
(cf. Eq.\rf{A-eqs}) and the canonical Hamiltonian becomes:
\be
\cH = -\frac{1}{\sqrt{-\g} \g^{00}} \h \Bigl\lb \frac{1}{E} \cP^2 +
E \(\pa_\s X\)^2 + \frac{1}{E} \(\pi_v + \eps A_1\)^2 + E \(\pa_\s v\)^2
\Bigr\rb
+ \frac{\g^{01}}{\g^{00}}\Bigl\lb \cP_\m \pa_\s X^\m +
\(\pi_v + \eps A_1\) \pa_\s v \Bigr\rb
\lab{Ham-canon-v}
\ee
We have skipped in \rf{Ham-canon-v} the linear combination of the rest of
the primary \rf{prim-constr} and secondary \rf{second-constr-a} constraints   
which remain unaltered by the presence of the new field $v$ except
for the ``Gauss law'' constraint which now reads (cf. \rf{momenta-1-v}) :
\be
\pa_\s \( E + \eps v\) = 0
\lab{Gauss-v}
\ee
One can check that the basic constraints entering \rf{Ham-canon-v}
span again a closed Poisson-bracket algebra which this time involves also the
``Gauss law'' constraint \rf{Gauss-v} and the following {\em variable}
string tension equal to the world-sheet ``electric'' field \rf{momenta-1-v} :
\be
T \equiv E = C - \eps v
\lab{vari-string-tension-v}
\ee
% %

\end{document}